\documentclass[english, 12pt]{article}
\usepackage[T1]{fontenc}
\usepackage[latin9]{inputenc}
\usepackage{geometry}
\usepackage{float}
\usepackage{graphicx,mathrsfs}
\usepackage{babel,soul}
\usepackage[makeroom]{cancel}

\def\Wbb{\mathbb{W}}

\usepackage{amsfonts}
\usepackage{amsmath}
\usepackage{amssymb}
\usepackage{amsthm}
\usepackage{bm}
\usepackage{natbib}
\usepackage{hyperref}
\usepackage{xcolor}
\usepackage{tcolorbox}
\usepackage{setspace}
\usepackage{authblk}

\usepackage{algorithm}
\usepackage{algpseudocode}

\DeclareMathOperator*{\argmax}{arg\,max}

\def\Op{O_{\sf P}}
\def\op{o_{\sf P}}
\def\bu{\boldsymbol{u}}

\def\trans{^{\scriptscriptstyle \sf T}}

\def\bX{\bm X}
\def\bx{\bm x}

\def\ba{\bm a}
\def\bSigma{\bm \Sigma}

\def\btheta{\bm\theta}

\def\bZ{\bm Z}

\def\bA{\bm A}
\def\bB{\bm B}
\def\bC{\bm C}

\def\bD{\bm D}
\def\bE{\bm E}

\def\bV{\bm V}
\def\bW{\bm W}

\def\bU{\bm U}

\def\bM{\bm M}
\def\bs{\bm s}
\def\bS{\bm S}
\def\bU{\bm U}

\def\bone{\bm 1}

\def\bBhat{\widehat\bB}

\def\bLambda{\bm \Lambda}
\def\bLambdahat{\widehat\bLambda}
\def\bOmega{\Omega}

\def\bgamma{\bm \gamma}

\def\bzeta{\bm\zeta}
\def\bxi{\bm\xi}

\def\trans{^{\scriptscriptstyle \sf T}}
\def\szero{^{\scriptscriptstyle \sf (0)}}
\def\sone{^{\scriptscriptstyle \sf (1)}}
\def\sy{^{\scriptscriptstyle \sf (y)}}
\def\sYi{^{\scriptscriptstyle (Y_i)}}

\def\supyzero{^{\scriptscriptstyle \sf (y)[0]}}

\def\supzero{^{\scriptscriptstyle \sf [0]}}

\def\subnew{_{\sf new}}

\def\subt{_{[t]}}
\def\subzero{_{[0]}}
\def\supzero{^{[0]}}
\def\supyzero{^{(y),[0]}}
\def\subtpone{_{[t+1]}}

\def\ba{\boldsymbol{a}}
\def\expit{{\rm expit}}
\def\boldb{\boldsymbol{b}}
\def\Err{{\rm Err}}

\definecolor{darkred}{RGB}{150,50,50}
\definecolor{brown}{RGB}{250,100,100}
\definecolor{green}{RGB}{000,150,100}
\definecolor{purple}{RGB}{250,000,180}

\newtheorem{lemma}{Lemma}

\newtheorem{theorem}{Theorem}
\newtheorem{remark}{Remark}
\newtheorem{assumption}{Assumption}
\newtheorem{corollary}{Corollary}
\newtheorem{proposition}{Proposition}

\newcommand{\indep}{\perp \!\!\! \perp}

\begin{document}

\title{Semi-supervised Clustering Through Representation Learning of Large-scale EHR Data}

\author[1]{\normalsize Linshanshan Wang$^*$}
\author[2]{Mengyan Li\footnote{Linshanshan Wang and Mengyan Li contributed equally to this work.}}
\author[3]{Zongqi Xia}
\author[4,5]{Molei Liu\footnote{Correspondence: moleiliu@bjmu.edu.cn.}}
\author[1]{Tianxi Cai}

\affil[1]{Department of Biostatistics, Harvard T.H. Chan School of Public Health} 
\affil[2]{Department of Mathematical Sciences, Bentley University} 
\affil[3]{Department of Neurology, University of Pittsburgh} 
\affil[4]{Department of Biostatistics, Peking University Health Science Center} 
\affil[5]{Beijing International Center for Mathematical Research, Peking University} 

\maketitle

\begin{abstract}

\noindent Electronic Health Records (EHR) offer rich real-world data for personalized medicine, providing insights into disease progression, treatment responses, and patient outcomes. However, their sparsity, heterogeneity, and high dimensionality make them difficult to model, while the lack of standardized ground truth further complicates predictive modeling. To address these challenges, we propose SCORE, a semi-supervised representation learning framework that captures multi-domain disease profiles through patient embeddings. 
SCORE employs a {\bf P}oisson-{\bf A}dapted {\bf L}atent factor {\bf M}ixture (PALM) Model with pre-trained code embeddings to characterize codified features and extract meaningful patient phenotypes and embeddings. To handle the computational challenges of large-scale data, it introduces a hybrid Expectation-Maximization (EM) and Gaussian Variational Approximation (GVA) algorithm, leveraging limited labeled data to refine estimates on a vast pool of unlabeled samples. 
We theoretically establish the convergence of this hybrid approach, quantify GVA errors, and derive SCORE's error rate under diverging embedding dimensions. Our analysis shows that incorporating unlabeled data enhances accuracy and reduces sensitivity to label scarcity. Extensive simulations confirm SCORE's superior finite-sample performance over existing methods. Finally, we apply SCORE to predict disability status for patients with multiple sclerosis (MS) using partially labeled EHR data, demonstrating that it produces more informative and predictive patient embeddings for multiple MS-related conditions compared to existing approaches.

\end{abstract}


\section{Introduction}
\subsection{Background}

{
Advances in electronic health records (EHR) have enabled unprecedented opportunities for leveraging large-scale, real-world clinical data to improve disease understanding, diagnosis, and management. It has become an important source for a growing number of data-driven biomedical studies that aim to extract clinical knowledge \citep{hong2021clinical,li2024multisource}, evaluate treatment response \citep{zhou2023multi}, and predict disease progression to guide clinical practice \citep{landi2020deep}. However, despite its promise, EHR data poses significant challenges for modeling that limit its utility in research and clinical applications. One of the main challenges arises from the high dimensionality of the features.

To address this, modern machine learning techniques have been developed to handle high-dimensional data, with \textit{representation learning} emerging as a particularly successful paradigm. Originally pioneered in fields such as nature language processing and image recognition \citep{bengio2013representation,devlin2019bert,payandeh2023deep}. Representation learning aims to project the high dimensional features into a lower dimensional space that retains the most relevant information for feature detection or classification tasks. Compared with traditional supervised machine learning methods, such models can generalize better to unseen data or related but different outcomes, leading to more robust and accurate predictions \cite{learmonth2013validation}. This is especially helpful in the EHR study setting given the inherently multi-domain and multi-outcome nature of many diseases, where different clinical metrics and outcomes often reflect overlapping yet distinct aspects of the disease. For example, in patients with multiple sclerosis (MS), disability progression is often assessed using scores such as the Expanded Disability Status Scale (EDSS) \citep{kurtzke1983rating} and the Patient-Determined Disease Steps (PDDS) \citep{kister2013disability}. While both scales aim to capture disability severity, EDSS provides a clinician-assessed, detailed evaluation across multiple functional systems, whereas PDDS is a patient-reported outcome measure that offers a simpler, subjective perspective. These scores, though correlated, reflect overlapping yet differentiated facets of disease progression \citep{foong2024patient}. Moreover, learned representations can be transferred to new tasks such as prediction of future outcomes, reducing the need for extensive training on new datasets.

However, despite its success in many other domains, representation learning with theoretical guarantees for patient clustering in the EHR setting remains a significant challenge -- particularly when aiming to improve both downstream prediction and patient subtyping -- due to limited sample sizes and the scarcity of labeled data, especially in the context of rare diseases like multiple sclerosis. Nonetheless, pre-trained representations of EHR feature derived from large-scale generic EHR data \citep{zhou2023multi} and semantic representations of EHR features \citep{lee2020biobert, rasmy2021med, yuan2022coder} provide an opportunity to guide model training effectively. Furthermore, even with a small number of labeled samples, representation learning can capture nuanced aspects of each patient's health profile, leading to richer, more informative embeddings and predictions than conventional methods. By leveraging existing knowledge of EHR feature representations and incorporating task-specific labeled data, representation learning holds promise for enhancing patient classification and predictive modeling in real-world clinical applications.

\subsection{Related Work}

Multivariate count data arise in diverse fields such as text mining, ecology, genomics, and EHR analysis, where capturing dependencies among variables is essential for meaningful interpretation. 
Embedding techniques, used to map complex data into lower-dimensional spaces while preserving essential structure, have become central to modern machine learning, powering advances in language modeling, sentiment analysis, and medical informatics \citep{mikolov2013efficient, devlin2018bert}. In the context of EHR data, for instance, Med-BERT leverages pretraining tailored to EHRs for better outcome prediction \citep{rasmy2021med}, while unsupervised deep models extract patient embeddings that enhance clinical inference \citep[e.g.,][]{miotto2016deep}.
Despite their scalability, these unsupervised methods lack theoretical guarantees and may be too generic for specific predictive tasks. To address this, we propose to use these pretrained embeddings as prior knowledge, and further fine-tune them on task-specific patient cohorts with partially observed labels. 

Beyond deep learning, classical statistical methods such as factor analysis and probabilistic PCA have long been used to uncover latent structures in data \citep{harman1976modern, tipping1999probabilistic}. However, their reliance on Gaussian assumptions often makes them inadequate for modeling real-world count data, such as medication usage patterns in EHRs. 
Generalized factor models address this limitation by introducing latent variables and applying maximum likelihood estimation (MLE) via Expectation-Maximization (EM) to handle non-normality \citep{moustaki2000generalized}.
For count data, the multivariate Poisson-lognormal (PLN) model \citep{aitchison1989multivariate} provides a hierarchical framework that incorporates multivariate Gaussian latent variables to effectively capture dependencies among counts. 
However, such methods face scalability challenges in high-dimensional settings due to intractable multi-dimensional integration over latent factors. 
Alternative quadrature methods \citep{schilling2005high} and Bayesian approaches such as Markov Chain Monte Carlo (MCMC) sampling \citep{edwards2010markov} offer greater modeling flexibility but often become computationally prohibitive, particularly for large EHR datasets involving numerous latent factors.

To circumvent the computational burden of high-dimensional integration, some approaches treat latent factors as fixed parameters and estimate them via singular value decomposition (SVD) \citep{zhang2020note} or constrained joint MLE \citep{chen2019joint}. While computationally efficient, these methods lack the regularization properties of random-effects models, increasing the risk of overfitting and instability.
Additionally, by ignoring uncertainty in the factor estimates, they are less appropriate in settings characterized by population heterogeneity and are unable to capture latent class mixtures, an essential limitation for applications such as patient clustering and sub-phenotyping.
They may also face potential identifiability issues when incorporating covariates.

These challenges underscore the need for probabilistic embedding frameworks that balance computational efficiency and interpretability, while effectively capturing uncertainty and structured relationships in non-normal data.
One prominent class of approaches addresses this need by modeling the posterior distributions of latent factors, rather than relying solely on point estimates. 
For instance, \citet{huber2004estimation} apply a Laplace approximation to the likelihood function and model the posterior distribution of latent factors as Gaussian.
Variational inference (VI) \citep{jordan1999introduction, wainwright2008graphical} offers a more flexible alternative by approximating the true posterior with a distribution from a pre-specified family, selected to minimize the Kullback-Leibler (KL) divergence. 
The choice of the approximation family involves a trade-off between flexibility in capturing posterior complexity and computational efficiency \citep{blei2017variational}. Among available methods, Gaussian variational approximation (GVA) \citep{kucukelbir2015automatic, opper2009variational} is widely used due to its analytical tractability and scalability. GVA approximates the posterior with a Gaussian distribution, optimizing its mean and covariance to minimize KL divergence. Compared to quadrature method and MCMC, GVA is computationally more efficient and scales better to large datasets \citep{blei2017variational, kucukelbir2015automatic}. 

GVA has been successfully applied to PLN models to mitigate computational costs in large-scale count-based data \citep{chiquet2018variational, chiquet2019variational}. 
Recent advancements in PLN methods have further expanded their applicability by incorporating covariates and offsets, enabling tasks such as multivariate regression, linear discriminant analysis, model-based clustering, and network inference \citep{chiquet2018variational, chiquet2019variational, silva2019multivariate}.
However, these models, which lack theoretical guarantees, do not readily extend to semi-supervised settings with limited labeled data. They also lack mechanisms to adaptively incorporate external pretrained embeddings for reducing parameter complexity and enhancing estimation efficiency.

Despite the empirical success of VI and GVA, the theoretical foundations remain largely unexplored, 
raising key questions about how approximation error scales with the dimensionality of latent factors, particularly in complex models like PLN \citep{blei2017variational}. Most existing theoretical results focus on simpler settings, such as Gaussian mixture models \citep{titterington2006convergence}, stochastic block models \citep{celisse2012consistency, bickel2013asymptotic}, and Bayesian linear models \citep{you2014variational, ormerod2017variational}. More relevant to our setting, \cite{hall2011theory, hall2011asymptotic} studied a Poisson mixed-effects model with a single predictor and repeated outcome measurements, a special case of PLN with rank one, and established parametric convergence and asymptotic normality of GVA. 
However, these results do not extend to the more complex scenario we consider, which involves a low-rank structure with diverging latent dimensionality, latent class membership, and semi-supervised data. 
In particular, the presence of latent clusters among unlabeled subjects prevents GVA from accurately approximating the full posterior distribution; instead, the posterior is better captured by a mixture of GVAs induced within the EM algorithm as proposed in our framework. This observation underscores the need for a rigorous theoretical framework to assess the reliability of GVA-based inference, including mixtures of GVAs employed within EM, for high-dimensional count data, especially under structured dependencies and semi-supervised settings.

\subsection{Our Contribution}

We propose {\bf S}emi-supervised {\bf C}lustering thr{\bf O}ugh {\bf RE}presentation learning (SCORE), a unified framework designed to advance patient clustering and embedding construction in the presence of label scarcity, population heterogeneity, and complex, multi-domain disease landscapes.
SCORE extends PLN-based clustering approaches and bridges a key methodological and theoretical gap by integrating semi-supervised learning, representation learning, GVA-based inference, and external pre-trained embeddings of EHR codes.
The framework improves scalability, estimation efficiency, robustness to label noise, adaptivity to knowledge transfer, and interpretability in high-dimensional semi-supervised settings, thereby enabling more accurate patient stratification and disease sub-phenotyping, particularly when labeled data are limited. Our contributions span three key aspects listed below.

\vspace{-0.3cm}

\paragraph{Powerful and flexible statistical learning framework.} In EHR settings, the dimensionality of observed variables (e.g., EHR codes) and patient embeddings can be large relative to the target cohort size, particularly for rare diseases, posing challenges in estimating the loading parameter matrix, which represents EHR code embeddings. To address this, SCORE introduces embedding adaptation as a transfer learning strategy, leveraging pre-trained EHR code embeddings from larger datasets with lower estimation errors. Rather than using these code embeddings directly, SCORE retains their linear structure while estimating only the scaling parameters (singular values) in the target cohort, significantly reducing the number of parameters to estimate while allowing for adaptation to target-specific heterogeneity. This approach improves computational efficiency and minimizes estimation error. SCORE integrates this code representation adaptation into a 
{\bf P}oisson-{\bf A}dapted {\bf L}atent factor {\bf M}ixture (PALM) modeling framework, refining pre-trained embeddings with the target data to enhance representation expressiveness while adjusting for baseline covariates and partially observed labels. Unlike traditional clustering and classification methods that focus on single-outcome modeling, SCORE learns latent patient embeddings that generalize across related clinical outcomes, enabling a robust and transferable characterization of disease progression.

To address computational bottlenecks in large-scale data, SCORE introduces a hybrid computational framework that combines GVA with EM. Distinct from prior work, SCORE approximates the posterior distribution using a mixture of GVAs conditional on the latent cluster memberships. 
Critically, SCORE leverages the supervised estimator from the small labeled dataset to initialize EM, ensuring stable convergence even when large amounts of unlabeled data dominate the training objective, making it non-convex. This semi-supervised learning approach mitigates sensitivity to label scarcity, accelerates training, and enhances the model's ability to learn generalizable patient representations from incomplete data.

\vspace{-0.3cm}

\paragraph{Theoretical advancements.} We provide a rigorous theoretical foundation for our SCORE algorithm, establishing the convergence of the hybrid EM-GVA algorithm under diverging embedding dimensions and semi-supervised learning regimes. Specifically, we quantify the error of the mixture of GVAs and characterize its dependence on the dimensionality of codified features and latent embeddings, addressing a critical gap in variational inference for low-rank Poisson mixture models, where such analysis has been largely unexplored. Furthermore, we demonstrate that our EM-GVA algorithm exhibits linear contraction, leveraging a supervised estimator for initialization to ensure stable convergence, even when the unlabeled data dominates the training objective. Importantly, we show that the resulting semi-supervised estimator is unaffected by label scarcity and achieves a significantly faster convergence rate than its supervised counterpart. Our theoretical analysis extends beyond existing work, tackling the unique challenges posed by high-dimensional empirical processes under diverging dimensionality and the hybrid nature of our optimization framework, which integrates variational approximation with iterative refinement via EM.

\vspace{-0.3cm}

\paragraph{Empirical validation.}
Extensive simulations demonstrate the superiority of SCORE in parameter estimation, embedding quality, and prediction accuracy over existing methods. Applied to MS phenotyping using real-world EHR data, SCORE achieves better classification performance and improved generalizability across multiple disability metrics compared to existing methods. In addition, trained patient embeddings are highly effective in predicting future outcomes. Our framework's ability to leverage unlabeled data also enhances robustness to the scarcity and noisiness of the labels compared to supervised learning in practice. 

\section{Methods}

\def\bUvec{\vec{\bU}}
\def\bUvecY{\bUvec_{Y}}
\def\bUvecYi{\bUvec_{Yi}}
\def\Rbb{\mathbb{R}}
\def\Dscr{\mathscr{D}}
\def\Isc{\mathcal{I}}
\subsection{Model Setup and Estimation Challenges}\label{sec:model:setup}
Let $\bX = (X_1, \dots, X_p)\trans$ represent the observed count vector of EHR concepts, $\bU$ denote the $r$-dimensional baseline covariate vector, including demographic information and healthcare utilization, and $Y$ be the binary class label. We assume that $\bX$ and $\bU$ are observed for all patients, while $Y$ is available only for a small labeled subset. Let $\Dscr = \{(\bX_i\trans, \bU_i\trans, \delta_i Y_i)\trans, i \in [N]\}$ denote the observed $N$ independent and identically distributed realizations of $(\bX\trans, \bU\trans, \delta Y)\trans$, where $\delta_i = I(i \in \Isc_L)$ indicates whether $Y_i$ is observed, and $\Isc_L \subset [N] = \{1, \dots, N\}$ indexes the labeled set.
We assume that the labeling process is random but can depend on $\bU_i$. 
We may also write $\Dscr = \Dscr_L \cup \Dscr_U$, where $\Dscr_L = \{(\bX_i\trans,\bU_i\trans,Y_i)\trans,i\in\Isc_L\}$, $\Dscr_U = \{(\bX_j\trans,\bU_j\trans)\trans, j \in \Isc_U\}$, and $\Isc_U= [N] \setminus \Isc_L$. 
Without loss of generality, we assume the first $n$ observations are labeled. 
We introduce $\bV \in \mathbb{R}^{p \times q}$ to denote the linear subspace spanned by the pre-trained embeddings of the $p$ EHR concepts, thereby encoding prior knowledge about the latent structure of $\bX$. For identifiability, we assume that $\sqrt{q/p} \bV$ is orthonormal.

We consider a {\bf P}oisson-{\bf A}dapted {\bf L}atent factor {\bf M}ixture (PALM) Model:
\begin{equation}\label{eq:model}
\begin{aligned}
&\bX\mid \bZ \sim \text{Poisson}(e^{\bZ}), \quad X_{1} \perp X_2 \perp \cdots \perp X_p \mid \bZ \\
&\bZ   =  \bV\left(\bB \bUvecY + \bW \right), \quad \bUvec_{Y} = (1, \bU\trans, Y)\trans \in \Rbb^{r+2}, \quad \bW\sim \mathcal N(\bm 0_{q}, \bLambda),  \\
& Y \mid \bU \sim \text{Bernoulli}\{ \expit(\boldb\trans \bUvec)\}, \quad \expit(v)=e^v/(1+e^v), \quad \bUvec = (1,\bU\trans)\trans. \\
\end{aligned}
\end{equation} 
This latent factor model facilitates embedding adaptation, enabling knowledge transfer from $\bV$ to efficiently learn latent patient representations. The random effect $\bW$ captures individual variability. The coefficient matrix $\bB = (\bB_1,\bB_U,\bB_Y) \in (\mathbb{R}^q, \mathbb{R}^{q\times r}, \mathbb{R}^q)$ encodes the intercept ($\bB_1$), the effect of the baseline covariates $\bU$ ($\bB_U$), and the effect of $Y$ ($\bB_Y$) on $\bX$ in the log scale. We consider a regime that the dimension of the observed EHR codes $p$, the dimension of the latent space $q$, and the total sample size $N$ all increase and diverge alone with the labeled sample size $n$ while the number of baseline covariates $r$ is fixed.
We further assume $q\ll p$ and $n = |\Isc_L| \ll N$ in our semi-supervised regime.

The PALM defined in (\ref{eq:model}) represents each subject's latent embedding as $\bar\bxi: =\bB \bUvecY + \bW \in \mathbb{R}^q$, where $\bar{\boldsymbol{\xi}}$ follows a multivariate normal distribution with mean $\bB \bUvecY$ and covariance matrix $\bLambda$, effectively capturing both structured effects and individual variability. Under (\ref{eq:model}), we have
\[
Y\indep \bX \mid(\bar\bxi,\bU),
\]
indicating that for some unlabeled subject with observed $(\bX,\bU)$, we will not lose any information about the underlying $Y$ by replacing the high-dimensional $(\bX, \bU)$ with the low-dimensional $(\bar\bxi,\bU)$. This property is rigorously established in Proposition \ref{prop:1} that justifies the cross-entropy loss of $Y$ against $\Pr(Y=1| \bar\bxi, \bU)$ is not larger than that of $\Pr(Y=1| \bX, \bU)$.

\begin{proposition}
\label{prop:1}
Under the PALM (\ref{eq:model}), 
\begin{align*}
&-\mathbb{E}\big[I(Y=1)\log \Pr(Y=1| \bar\bxi, \bU)+I(Y=0)\log \Pr(Y=0| \bar\bxi, \bU)\big]\\
\leq &-\mathbb{E}\big[I(Y=1)\log \Pr(Y=1| \bX, \bU)+I(Y=0)\log \Pr(Y=0| \bX, \bU)\big].
\end{align*}
\end{proposition}

Moreover, the patient embedding $\bar\bxi$, summarizing the high-dimensional EHR count data, can serve as compact representations for phenotypes related to, but not identical to, $Y$, and thus, can be utilized in more comprehensive downstream tasks. 
For example, in Section \ref{sec:realdata}, we demonstrate that patient embeddings trained using PDDS as the target phenotype $Y$ are highly predictive of another key phenotype, the EDSS-defined disability status.
This illustrates the ability of the learned representations to generalize across related phenotypes and support transferable characterization of complex disease progression.

\begin{remark}

Pre-trained EHR code embeddings from large language models or large cohorts \citep[e.g.]{hong2021clinical, xiong2023knowledge} offer us valuable prior knowledge and potential gain of both computational and statistical efficiency. This is because we can reduce the number of parameters from $O(pq)$ to $O(q^2)$ by bypassing the estimation of $\bV$. However, directly transferring some pre-trained embeddings may fail to capture the heterogeneity between the external source and the target population and the disease of our interest. To address this issue in model (\ref{eq:model}), we do not set the covariance matrix of $\bW$ as identity, which is a conventional setup of the latent factor analysis with a totally unknown loading matrix used to ensure identifiability. Instead, we allow the covariance parameter $\bLambda$ to be freely driven by the target population. In this way, we only leverage the linear subspace of the external code embeddings and adjust for the scale heterogeneity as well as the rotation ambiguity between the external source and the target.
\end{remark}

A natural approach to estimating the unknown model parameters $\btheta = \{\bB, \bLambda, \boldb\}$ is maximum likelihood estimation (MLE). If $Y_i$ and $\bW_i$ were observed, then the complete-data log-likelihood can be written as  
\begin{equation}\label{eq:loglik}
\begin{aligned}
   &\sum_{i \in [N] } \log P(\bX_i, \bW_i, Y_i | \bU_i; \btheta) \\
   = &\sum_{i \in  [N] } 
 \left\{\log P(\bX_i| \bW_i, \bU_i, Y_i; \bB) + \log P(\bW_i;\bLambda) + \log P(Y_i | \bU_i;\boldb)\right\},
\end{aligned}
\end{equation}
where we use notation $P(\bA_1,...,\bA_k \mid \bC_1,...,\bC_l; \Theta)$ the denote the conditional density of $\bA_1, ..., \bA_k$ given $\bC_1, ..., \bC_l$ and the parameters $\Theta$. Since $\bW_i$ is latent and $Y_i$ is not observed for $i \in \Isc_U$, the MLE has to be constructed based on the log-likelihood of the observed data, 
$\ell(\btheta) = \ell_{L}(\btheta) + \ell_{U}(\btheta)$,
where 
\begin{equation*}
\begin{split}
\ell_{L}(\btheta) = &\sum_{i \in \mathcal I_{L}}  \log P(\bX_i, Y_i| \bU_i; \btheta) = \sum_{i \in \mathcal I_{L}}  \log \int  P(\bX_i, \bW_i, Y_i | \bU_i; \btheta) d\bW_i;\\
\ell_{U}(\btheta) =& \sum_{i \in \mathcal 
 I_{U}} \log\{ P(\bX_i, Y_i=0| \bU_i; \btheta) +  P(\bX_i, Y_i=1| \bU_i; \btheta)  \} \\
 =& \sum_{i \in \mathcal 
 I_{U}}  \log \sum_{y=0, 1}\int  P(\bX_i, \bW_i, Y_i=y | \bU_i; \btheta) d\bW_i.      
\end{split}   
\end{equation*}
The objective function $\ell(\btheta)$ does not have a closed-form expression due to the integration over $\bW_i$, rendering exact maximization with respect to $\btheta$ computationally intractable. To address this, prior work, including \cite{karlis2005algorithm} and \cite{silva2019multivariate}, has explored approximating the log-likelihood via numerical or Monte Carlo integration. However, these approaches quickly become computationally prohibitive, even for moderate embedding dimensions $q$, highlighting the need for additional approximation strategies to maintain scalability.

\subsection{Gaussian Variational Approximation (GVA) Strategy}\label{sec:method:gva}
To address the computational challenges posed by the integral over $\bW_i$, we adopt a variational approximation approach. Specifically, we approximate the intractable posterior distribution $P(\bW_i|\bX_i, \bU_i, Y_i = y)$ with a {\em subject-specific} and tractable distribution, denoted by $\phi_i^{(y)}(\bW_i)$. This local approximation is tailored to each individual observation. For the $i$-th subject, it enables us to approximate the log-likelihood term $\log {P(\bX_i, Y_i = y| \bU_i; \btheta)}$ using
$\mathcal{J}_i\sy= \mathcal J(\btheta, \phi_i\sy; \bX_i,\bU_i, y)$, where
$$
\begin{aligned} 
&\mathcal J(\btheta, \phi\sy; \bx,\bu, y)\\
:=&\log \{P(\bX = \bx, Y =y | \bU = \bu; \btheta)\} - {\rm KL}\{\phi\sy(\bW), P(\bW|\bX = \bx,\bU = \bu, Y= y; \btheta)\} \\
=&\mathbb E_{\bW \sim \phi\sy}\log \{P(\bX = \bx, \bW, Y = y| \bU = \bu; \btheta)\} - \mathbb E_{\bW \sim \phi\sy}\log \{\phi\sy(\bW)\},
\end{aligned}
$$
where ${\rm KL}(\cdot,\cdot)$ represents the KL-divergence between two distributions and $\mathbb E _{\bW \sim \phi\sy}$ denotes the expectation taken with respect to the conditional distribution defined by ${\phi\sy}(\bW)$. Since the KL term is nonnegative, we have $\mathcal{J}_i\sy\leq \log P(\bX_i, Y_i = y|\bU_i;\btheta)$ and thus refer to $\mathcal{J}_i\sy$ as an evidence lower bound (ELBO) of the log-likelihood function.
The key idea is to properly specify $\phi_i\sy$ such that: (i) $\mathcal{J}_i\sy$ admits a closed-form expression and can be computed without numerical integration; and (ii) the KL-divergence between $\phi_i\sy$ and the true posterior distribution of $\bW_i$ can potentially vanish. In this way, we can maximize the ELBO as a surrogate for $\log P(\bX_i, Y_i = y | \bU_i; \btheta)$, substantially reducing the computational burden with minimal loss in statistical accuracy and efficiency.

\def\Wbb{\mathbb{W}}

For the construction of each $\phi_i\sy$, we define the set of candidate GVA densities as 
\begin{align*}
\Big\{\phi_i\sy(\bW_i) =\mathcal N(\bW_i; \bm m_i\sy, \bm S_i\sy):~\bm m_i\sy\in \mathbb R^q,~\bm S_i\sy = {\rm diag}(\bm s_i\sy),~\bs_i\sy \in\mathbb R^q_+, ~y=0,1\Big\},    
\end{align*}
where
$\mathcal N(\cdot; \bm m_i\sy, \bm S_i\sy)$ denotes the probability density function of the multivariate Gaussian distribution with mean $\bm m_i\sy$ and covariance $\bm S_i\sy$, ${\rm diag}(\mathbf{a})={\rm diag}(a_1,...,a_L) \in \mathbb{R}^{L\times L}$ for any vector $\mathbf{a}=(a_1,...,a_L)$, and $\mathbb R^q_+$ denotes the set of $q$-dimensional vectors with all positive elements. Utilizing the Gaussian density formation of $\phi_i\sy$, we derive the explicit form of $\mathcal{J}_i\sy$:
\begin{equation}\label{eq:elbo_i}
    \begin{aligned}
     \mathcal{J}_i\sy =& \sum_{j=1}^p X_{ij} \bV_j\trans (\bB \bUvec_{iy}  + \bm{m}_i\sy ) -\sum_{j=1}^p A_{ij}\sy - \sum_{j=1}^p \log (X_{ij}\!)\\
    &- \frac{1}{2} \log \{{\rm det}(\bLambda)\} - \frac{1}{2} (\bm{m}_i\sy)\trans \bLambda^{-1}\bm{m}_i\sy - \frac{1}{2}{\rm Tr}(\bLambda^{-1}\bS_i\sy) + \frac{1}{2} \log \{{\rm det}(\bS_i\sy)\} +\frac{q}{2}  \\
    & + y\log\{\expit(\boldb\trans\bUvec_i)\} + (1-y) \log\{1- \expit(\boldb\trans\bUvec_i)\},
\end{aligned}
\end{equation}
with the definition
\begin{equation*}
    A_{ij}\sy:= \mathbb E_{\bW \sim \phi_i\sy}\exp\{ \bV_j\trans (\bB \bUvec_{iy} + \bW_i)\} 
    = \exp\Big\{ \bV_j\trans\Big (\bB  \bUvec_{iy} + \bm{m}_i\sy + \frac{1}{2} \bS_i\sy\bV_j\Big)\Big\},
\end{equation*}
$\bUvec_{iy}  = (1, \bU_i\trans, y)\trans$, and ${\rm det} (\bA)$ being the determinant of matrix $\bA$.
Since $\phi_i\sy$ is parameterized by $\bzeta\sy_i:=\{\bm{m}_i\sy,\bS_i\sy\}$, we also use
$\mathcal J(\btheta,\bzeta\sy; \bx, \bu, y)$ to denote the ELBO for the specific choice of $\phi\sy$ and let $\mathcal{J}_i\sy=\mathcal J(\btheta,\bzeta\sy_i; \bX_i, \bU_i, y)$ correspondingly. The tractable form of $\mathcal{J}_i\sy$ presented in (\ref{eq:elbo_i}) will be used to construct the objective function for $\btheta$ replacing the intractable log-likelihood of the observed data, which can help remove time-consuming numerical integration and significantly reduce computational complexity. This is crucial in large-scale biomedical studies where $N$, $p$, and $q$ are typically large.

The above introduced variational approximation strategy is three-fold. First, we select the Gaussian family for its analytical advantage within the PALM framework, enabling an explicit expression for $\mathcal J_i\sy$. 
Second, we constrain the variational covariance $\bS_i\sy$ to a diagonal matrix, avoiding the computational instability and the slowness of inverting a potentially large matrix iteratively. Third, we adopt the mean-field variational family, assuming mutual independence of the latent variable $\bW_i$ between different observations conditional on observed data. In Section \ref{sec:theory}, we carefully study and establish the approximation error rate of the GVA under reasonable regularity conditions.

\subsection{Estimation Strategy}
In the SSL setting, we first apply GVA to the labeled data to obtain an initial estimator $\widetilde{\btheta}$. This preliminary estimate is then refined using a hybrid approach that combines GVA with a latent-class EM algorithm, incorporating both labeled and unlabeled observations to produce the final SCORE estimator $\widehat{\btheta}$.

\subsubsection{Supervised Initial Estimator: Restricted Maximum ELBO}\label{method:est:sup}
Let  $\bM_{\mathcal{I}_L}=\{\bm m_i\sYi:~i\in \mathcal{I}_L\}$, $\bS_{\mathcal{I}_L}=\{\bS_i\sYi:~i\in \mathcal{I}_L\}$ and $\bzeta_{\mathcal{I}_L}=\{\bM_{\mathcal{I}_L},\bS_{\mathcal{I}_L}\}$ denote the GVA parameters for the labeled subjects. 
Note that for $i\in \mathcal{I}_L$, we only need to introduce and estimate the GVA parameter corresponding to the observed $Y_i$. As introduced in Section \ref{sec:method:gva}, instead of optimzing the intractable $\ell_L(\btheta)$, we consider maximizing the ELBO for $i\in \mathcal{I}_L$:
\begin{equation*}
\mathcal{J}_L(\btheta, \bzeta_{\mathcal{I}_L}; \Dscr_L) =\sum_{i\in \mathcal{I}_L} \mathcal{J}_i\sYi= \sum_{i\in \mathcal{I}_L}  \mathcal J(\btheta, \bzeta_i\sYi; \bX_i, \bU_i, Y_i).
\end{equation*}
Due to the diverging $q$ and the $\exp(\cdot)$ link in the PALM model, we need to introduce additional technical constraints to regularize the parameters. Specifically, we define that
\begin{equation}\label{eq:Omega_B}
\begin{split}
\Omega_B :=& \left\{\bB:~\|\bB\|_2\leq K_B\sqrt{q},~|\bV_j\trans \bB \bUvec_{iy} | \leq K_B \sqrt{\log (p/q)}, ~\forall i \in [N],~j \in[p]\right\},\\
\Omega_M^{L} :=& \left\{\bM_{\mathcal{I}_L}:~|\bV_j\trans \bm{m}_i\sYi| \leq K_M \sqrt{\log (p/q)},~ \forall i \in \mathcal{I}_L,~j \in[p]\right\},  
\end{split}
\end{equation}
where $K_B$ and $K_M$ are some positive finite constants. We propose to solve the following restricted maximum ELBO problem:
\begin{equation}\label{eq:opt}
 \widetilde\btheta,~\widetilde\bzeta_{\mathcal{I}_L}   = \arg \max_{\bB\in \Omega_B, \bM_{\mathcal{I}_L}\in \Omega_{M}^L}\mathcal{J}_L(\btheta, \bzeta_{\mathcal{I}_L}; \Dscr_L ),
\end{equation}
where $\widetilde\btheta$ is the supervised maximum ELBO estimator, serving as a preliminary estimate for the next step that incorporates all observations.
The rate $\sqrt{\log (p/q)}$ in (\ref{eq:Omega_B}) is derived from the high-probability $l_\infty$ upper bound for $p/q$-dimensional sub-Gaussian random vectors, where $p/q$ approximates the number of measurements for each factor/direction per observation. This sub-Gaussian upper bound is crucial for analyzing the convergence rate of our estimator. 
Unlike \cite{chen2019joint}, which assumes that the $\bW_i$'s are fixed and bounded, we consider slightly larger parameter spaces for the covariate-associated mean parameters and for the variational posterior means of individual observations.

From a profiling perspective, the GVA parameter of the $i$th observation given $Y_i = y$, denoted $\bzeta_{i}\sy$, can be viewed as a function of $(\bX_i, \bU_i, \btheta)$, satisfying
$$
\bzeta_{i}\sy = \bzeta\sy(\bX_i,\bU_i, \btheta); \quad \bzeta\sy(\bx,\bu, \btheta) = \arg \max_{\bzeta\sy}\mathcal J( \btheta, \bzeta\sy; \bx, \bu, y),
$$ 
with  $\bzeta\sy = \{{\bm{m}\sy}, {\bs\sy}\}$, subject to $|\bV_j\trans \bm{m}\sy| \leq K_M \sqrt{\log (p/q)}, ~\forall j \in[p]$.
Consequently, (\ref{eq:opt}) can be reformulated as 
$$
\widetilde \btheta = \arg\max_{\btheta} \sum_{i\in \mathcal{I}_L} \mathcal{Q}^{(Y_i)}( \bX_i, \bU_i, \btheta) \quad \text{subject to} \quad  \bB\in \Omega_B,
$$
where $\mathcal{Q}^{(Y_i)}( \bX_i, \bU_i, \btheta)$ is the optimal ELBO value with respect to the GVA parameters of subject $i$. Specifically, we have 
\begin{equation} \label{eq:profile}
    \mathcal{Q}\sy(\bx, \bu, \btheta) = 
\mathcal J\{ \btheta, \bzeta\sy(\bx, \bu, \btheta); \bx, \bu, y\}
\end{equation}
and it can be used to approximate $\log {P(\bX = \bx, Y = y \mid \bU= \bu;  \btheta)}$. 

We maximize the supervised ELBO by alternating between two steps until convergence: (1) gradient update on the GVA parameters $\bzeta_{\mathcal{I}_L}$ with $\btheta$ fixed; and (2) gradient update $\btheta$ keeping $\bzeta_{\mathcal{I}_L}$ fixed. Note that $\mathcal{J}_L(\btheta, \bzeta_{\mathcal{I}_L}; \Dscr_L )$ is strictly concave with respect to the model and GVA parameters, and features closed-form blockwise gradients. Based on this, we leverage the proximal gradient algorithm to update $\btheta$ and $\bzeta_{\mathcal{I}_L}$ subject to the constraints $\bB\in \Omega_B$ and $\bM_{\mathcal{I}_L}\in \Omega_{M}^L$. 

\subsubsection{SCORE Estimator: a Hybrid EM-GVA Algorithm}

To incorporate unlabeled data in $\Dscr_U$ with unobserved $Y$, a common strategy is to maximize the corresponding log-likelihood via the EM algorithm, iterating between: (I) E-step: estimating
$\Pr(Y_i=y| \bX_i, \bU_i; \btheta)$ and using it to impute the empirical version of 
\[
\mathbb{E}_{P(Y_i|\bX_i, \bU_i; \btheta)}\log \{P(\bX_i, Y_i|\bU_i; \btheta)\};
\]
and (II) M-step: updating $\btheta$ by maximizing the imputed log-likelihood function. In both steps, we must handle the intractable integral over the latent variable $\bW_i$. A key distinction from the supervised ELBO approach in Section \ref{method:est:sup} is that, for each unlabeled subject $i$, we must estimate two sets of variational parameters, $\bzeta_i^{(0)}$ and $\bzeta_i^{(1)}$, corresponding to the two possible labels $Y_i = 0$ and $Y_i = 1$, respectively. We use $\bM_{\mathcal{I}_U}=\{\bm m_i\sy:~i\in \mathcal{I}_U,~y\in\{0,1\}\}$, $\bS_{\mathcal{I}_U}=\{\bS_i\sy:~i\in \mathcal{I}_U,~y\in\{0,1\}\}$ and $\bzeta_{\mathcal{I}_U}=\{\bM_{\mathcal{I}_U},\bS_{\mathcal{I}_U}\}$ to denote the GVA parameters for the unlabeled data. To address the challenge arising from the need for both EM and GVA, we propose a hybrid latent-class EM and GVA (EM-GVA) algorithm. Starting from an initial estimator $\{\btheta\subzero,{\bzeta_{\mathcal{I}_L}}\subzero, {\bzeta_{\mathcal{I}_U}}\subzero\}$, 
the EM-GVA algorithm iterates for $T$ times between the E-step and M-step detailed below, ultimately producing the final SCORE estimator $\widehat{\btheta}$, which can be used for membership prediction and patient embedding construction.

\def\supGVA{^{\scriptscriptstyle \sf GVA}}

\paragraph{E-step.}
To estimate $\Pr(Y_i=1| \bX_i, \bU_i; \btheta)$ for each unlabeled subject $i$, we note that 
\[
\Pr(Y_i=1| \bX_i, \bU_i; \btheta)=\frac{P(\bX_i, Y_i = 1|\bU_i;\btheta)}{P(\bX_i, Y_i = 1|\bU_i;\btheta)+P(\bX_i, Y_i = 0|\bU_i;\btheta)}.
\]
Similar to Section \ref{sec:method:gva}, we use the ELBO $\mathcal{J}_i\sy=\mathcal{J}(\btheta,\bzeta_i\sy; \bX_i, \bU_i, y)$ in (\ref{eq:elbo_i}) to approximate $\log\{P(\bX_i, Y_i = y|\bU_i;\btheta)\}$ with $\btheta$ and $\bzeta_i\sy$ plugged in as the estimator at the current iteration. 
Since $\bzeta_i\sy$ at the current iteration can be viewed as a function of $(\bX_i, \bU_i)$ and $\btheta$  at the current iteration, 
we approximate $\gamma(\bx,\bu; \btheta) := \Pr(Y_i = 1 | \bX_i=\bx, \bU_i=\bu;\btheta)$ using (\ref{eq:profile}) as
\begin{equation}
\label{eq:gamma}
\gamma\supGVA(\bx,\bu;\btheta) := \frac{\exp\{\mathcal{Q}\sone(\bx,\bu,\btheta)\}}{\exp\{\mathcal{Q}\sone(\bx,\bu,\btheta)\}+\exp\{\mathcal{Q}\szero(\bx,\bu,\btheta)\}}.
\end{equation}
For $i \in \mathcal{I}_U$, we use $\gamma\supGVA_i$ to denote  $\gamma\supGVA(\bX_i,\bU_i;\btheta)$ for simplicity, and let $\bgamma\supGVA = \{\gamma\supGVA_i:i\in\mathcal{I}_U\}$.

\paragraph{M-Step.}
Define the full-sample ELBO with $Y_i$ imputed by $\gamma_i\supGVA$ for unlabeled subjects as
$$\mathcal O_{F}(\btheta,\bzeta_{\mathcal{I}_L}, \bzeta_{\mathcal{I}_U}; \Dscr, \bgamma\supGVA):=\sum_{i\in \mathcal{I}_L} \mathcal J(\btheta,\bzeta_i\sYi; \bX_i, \bU_i, Y_i) + \sum_{i\in \mathcal{I}_U} \mathcal O(\btheta,\bzeta_i\szero, \bzeta_i\sone; \bX_i, \bU_i, \gamma_i\supGVA),
$$ 
where 
\[
\mathcal O(\btheta,\bzeta_i\szero, \bzeta_i\sone; \bX_i, \bU_i, \gamma_i\supGVA):= \gamma_i\supGVA \mathcal J(\btheta,\bzeta_i\sone; \bX_i, \bU_i, Y_i =1) + (1-\gamma_i\supGVA)\mathcal J(\btheta,\bzeta_i\szero; \bX_i, \bU_i, Y_i = 0).
\]
We then solve for the model and GVA parameters $\{\btheta,\bzeta_{\mathcal{I}_L}, \bzeta_{\mathcal{I}_U}\}$ via the latent-class-imputed ELBO optimization problem:
\[
\max_{\bB \in \bOmega_B,\bM_{\mathcal{I}_L} \in \bOmega_{M}^L,\bM_{\mathcal{I}_U} \in \bOmega_{M}^U}\mathcal O_{F}(\btheta,\bzeta_{\mathcal{I}_L}, \bzeta_{\mathcal{I}_U}; \Dscr,\bgamma\supGVA),
\]
where $\bOmega_{M}^U:= \big\{\bM_{\mathcal{I}_U}: |\bV_j\trans \bm{m}_i\sy| \leq K_M \sqrt{\log (p/q)},  
~\forall i \in \mathcal{I}_U,~y\in\{0, 1\},~j \in[p]\big\}$ is a constraint set for the GVA's mean parameters of unlabeled subjects. In the M-step, we alternately update $\bzeta_{\mathcal{I}_L}, \bzeta_{\mathcal{I}_U}$ and $\btheta$ using the proximal (blockwise) gradient algorithm.

The initial estimator plays a crucial role in the convergence performance of EM algorithm \citep{balakrishnan2017statistical}. In our case, this is even pronounced due to the divergence of latent dimensionality $q$ and the use of GVA. We propose to initialize the hybrid EM-GVA algorithm with the supervised estimator obtained in Section \ref{method:est:sup}, i.e., $\btheta\supzero=\widetilde\btheta$ and each $\bzeta_i\supyzero$ being the maximizer of $\mathcal J(\widetilde\btheta, \bzeta_i\sy; \bX_i, \bU_i, y)$ subject to $|\bV_j\trans \bm{m}_i\sy| \leq K_M \sqrt{\log (p/q)}, ~\forall j \in[p]$. This strategy helps ensure that the imputed labels $Y_i$ reflect the true underlying pattern of interest, such as the severity of disability, rather than arbitrary or spurious groupings. It also significantly improves the convergence speed of the EM-GVA algorithm. The theoretical impact of the initial supervised estimator $\widetilde{\btheta}$ on the final estimator $\widehat{\btheta}$ is rigorously analyzed in Section \ref{sec:theory}. In brief, the influence of $\widetilde{\btheta}$ diminishes with each EM iteration and ultimately vanishes. As a result, as long as $\widetilde{\btheta}$ is reasonably accurate (i.e., consistent), its potentially large error, due to limited labeled data, does not affect the convergence rate of $\widehat{\btheta}$.

\subsection{Phenotyping and Embedding for Future Subjects}

A key advantage of adopting a random effects model is that it allows us to use the probabilistic model to predict $Y$ and derive low-dimensional embeddings of $\bX$ for a future subject with observed counts $\bX=\bx\subnew$ and baseline covariates $\bU=\bu\subnew$ generated from the PALM (\ref{eq:model}). Inspired by the GVA approximation of $
\Pr(Y = 1 | \bX, \bU)$ defined in (\ref{eq:gamma}), we propose to estimate the probability function $\gamma\subnew = \Pr(Y = 1 | \bX = \bx\subnew, \bU = \bu\subnew)$ as  
\begin{equation}
\gamma\supGVA\subnew = \gamma\supGVA(\bx\subnew,\bu\subnew;\widehat\btheta) := \frac{\exp\{\mathcal{Q}\sone(\bx\subnew,\bu\subnew,\widehat\btheta)\}}{\exp\{\mathcal{Q}\sone(\bx\subnew,\bu\subnew,\widehat\btheta)\}+\exp\{\mathcal{Q}\szero(\bx\subnew,\bu\subnew,\widehat\btheta)\}},
\label{equ:def:pred}
\end{equation}
where $\widehat\btheta$ is our final SCORE estimator of $\btheta$, the optimal ELBO value $\mathcal{Q}\sy(\bx\subnew,\bu\subnew,\widehat\btheta) =  \mathcal J(\widehat \btheta, \widehat\bzeta_{\rm new}\sy; \bx_{\rm new}, \bu_{\rm new}, y)$ for $y\in\{0,1\}$, and
\[
\widehat\bzeta\subnew\sy = (\widehat{\bm m}\subnew\sy, \widehat{\bm{s}}\subnew\sy) = \argmax_{\zeta\subnew\sy:~ |\bV_j\trans \bm{m}\subnew\sy| \leq K_M \sqrt{\log (p/q)}} \mathcal J(\widehat \btheta,\bzeta\subnew\sy; \bx\subnew, \bu\subnew, y).
\]

In addition to the phenotyping probability $\gamma\supGVA\subnew$, 
SCORE also learns an embedding for the new patient that characterizes the complex disease progression and can serve as input features for related phenotyping and prediction tasks. 
Our derived $\gamma\supGVA\subnew$ and the GVA parameters $\widehat\bzeta\subnew\sy$ of the subject with $(\bx\subnew,\bu\subnew)$ can be naturally combined to characterize the underlying subject embedding $\bar\bxi\subnew := \bB_0(1,\bu\subnew, Y\subnew)\trans + \bW\subnew$, the clinical importance of which has been discussed in Section \ref{sec:model:setup}.  Specifically, we impute the unknown $Y\subnew$ in $\bar\bxi\subnew$ with the estimated probability $\gamma\supGVA\subnew$ and use the GVA mean parameter $\widehat{\bm m}\subnew\sy$ to approximate the latent $\bW\subnew$, which yields the estimated embedding
\begin{equation}
\widehat\bE\subnew:= (1-\gamma\supGVA\subnew)\widehat\bE\subnew\szero + \gamma\supGVA\subnew\widehat\bE\subnew\sone,\quad\mbox{where}\quad \widehat\bE\subnew\sy=\widehat \bB (1,\bu\subnew, y)\trans + \widehat{\bm m}\subnew\sy,~y=0,1.
\label{equ:Enew}
\end{equation}
The convergence properties of $\gamma\supGVA\subnew$ and $\widehat\bE\subnew$ are established in Corollary \ref{coro:pred}. Interestingly, we show that $\gamma\supGVA\subnew$ converges to $\Pr(Y=1| \bar\bxi=\bar\bxi\subnew,\bU=\bu\subnew)$, a more or at least equally informative predictor of disease status compared to its posterior probability $\Pr(Y=1| \bX=\bx\subnew, \bU=\bu\subnew)$, as shown in Proposition~\ref{prop:1}.

\section{Theoretical Analysis}
\label{sec:theory}

In this section, we present the theoretical convergence results for the proposed SCORE estimator, beginning with the necessary notation and assumptions. For any vector $\ba = (a_1, \ldots, a_p)\trans \in \mathbb{R}^p$, we denote by $\|\ba\|_{\infty} = \max_{1\leq j \leq p}|a_j|$, $\|\ba\|_r = \sum_{j=1}^p |a_j|^r$ for $r\in(0, 1]$, and $\|\ba\|_2 = \sqrt{\sum_{j=1}^p |a_j|^2}$ . For any matrix $\bA=[\ba_1,\ldots,\ba_{p_1}]\trans \in \mathbb{R}^{p_1 \times p_2}$ with $\ba_j\in \mathbb{R}^{p_2}$, we define the F-norm $\|\bA\|_{F}=\sqrt{\sum_{j=1}^{p_1}\|\ba_j\|_2^2}$, $\|\bA\|_{r,\infty}=\max_{j\in[p_1]}\|\ba_j\|_r$ for $r>0$, and use $\lambda_{\min}(\bA)$ and $\lambda_{\max}(\bA)$ to denote the smallest and largest singular value of $\bA$. For two sequences $a_n$ and $b_n$, we denote by $a_n<\infty$ if $a_n$ is bounded by some constant, $a_n=O(b_n)$ if $\lim_{n\rightarrow\infty}|a_n/b_n|<\infty$, $a_n=o(b_n)$ if $\lim_{n\rightarrow\infty}a_n/b_n=0$, and $a_n=\op(b_n)$ or $a_n=\Op(b_n)$ if $a_n=o(b_n)$ or $a_n=O(b_n)$ with a probability approaching $1$. 
We use $a_n \precsim b_n$
to denote $a_n \leq Cb_n$ for some constant $C>0$,
and use $a_n \asymp b_n$ to denote $C\leq a_n/b_n \leq C^{\prime}$ for
some constants $C, C^{\prime}>0$.

Let $\btheta_0=\{\bB_0=(\bB_{1,0},\bB_{U,0},\bB_{Y,0}), \bLambda_0,  
\boldb_0\}$ denote the true (population-level) value of the model parameters $\btheta=\{\bB,\bLambda,\boldb\}$. Define that 
\[
\Err(\btheta)=\max\big\{\|(\bB-\bB_0)\trans\|_{2,\infty},\|\bB_{Y,0}\trans\bLambda_0^{-1}(\bLambda - \bLambda_0)\|_2, \|\boldb-\boldb_0\|_2\big\},
\]
and the signal-to-noise ratio (SNR) level $\psi^2:=\bB_{Y,0}\trans\bLambda_0^{-1}\bB_{Y,0}$. We introduce function $\Err(\cdot)$ to measure the convergence error of the estimators. Here, we only consider one particular projection of $\bLambda - \bLambda_0$ but not the whole matrix since this projection $\bB_{Y,0}\trans\bLambda_0^{-1}(\bLambda - \bLambda_0)$ is already sufficient to characterize the error in the predicted probabilities and the estimated embeddings. The SNR parameter $\psi$ reflects the relative strength of the association between $Y_i$ and $\bX_i$ compared to the variation of the random effect $\bW_i$. It is important in characterizing the contraction of the EM algorithm as well as the error rate of the SCORE estimator. We consider the regime that the sample size and dimensionality parameters $N$, $p$ and $q$ all diverge alone with $n\rightarrow\infty$. In addition to the model assumptions introduced in Section \ref{sec:model:setup}, we also require the following regularity assumptions.

\begin{assumption}
(i) The baseline covariate $\bU$ belongs to a compact domain and has a continuous differentiable probability density function. (ii) It holds that $c^{-1}<\lambda_{\min}(\mathbb{E}[ \bUvec_{Y} \bUvec_{Y} \trans])\leq \lambda_{\max}(\mathbb{E}[\bUvec_{Y} \bUvec_{Y} \trans]) < c$ for some constant $c>0$.
\label{asu:1}
\end{assumption}

\begin{assumption}
The following assumptions hold: (i) $\sqrt{q/p}\bV$ is orthonormal and $\|\bV\|_{2,\infty} \leq C_V$, where $C_V$ is a positive constant. (ii) $\bB_0 \in  \Omega_B$ defined in (\ref{eq:Omega_B}). (iii) $0 <\kappa^{-1} \leq \lambda_{\min}(\bLambda_0)\leq \lambda_{\max}(\bLambda_0) \leq \kappa$, where $\kappa$ is a positive finite constant. (iv) $\|\boldb_0\|_{2}\leq C_{\pi}$, where $C_{\pi}$ is a positive finite constant. 

\label{asu:2}

\end{assumption}

\begin{assumption}
We treat $p/q$ as the benchmark dimensionality parameter, and assume that $n \asymp (p/q)^{\eta_1}$, $q \asymp (p/q)^{\eta_2}$, and $N \asymp (p/q)^{\eta_3}$ for some $\eta_1 >0$,  $\eta_2 >0$ and $\eta_3 >0$.
Also assume that for some arbitrarily small constant $\epsilon>0$, $q=o(p^{1/2-\epsilon}/\psi)$, $q=o(n/\psi^4)$, $\psi=o(\{p/q\}^{1/4-\epsilon})$, and $n=O(N)$, where $\psi^2=\bB_{Y,0}\trans\bLambda_0^{-1}\bB_{Y,0}$ is the SNR.

\label{asu:3}
\end{assumption}

\begin{remark}
Assumption \ref{asu:1} imposes standard regularity conditions on the baseline covariate $\bU$ in view of the classic theory for M-estimation \citep{van2000asymptotic}. The incoherence condition $\|\bV\|_{2,\infty}\leq C_V$ in Assumption \ref{asu:2} has been commonly used in the literature of matrix completion and factor analysis \citep[e.g.]{candes2011robust}.
It essentially asserts that the column vectors are not too `aligned' with the standard basis vectors, ensuring that no single entry of the matrix disproportionately influences the matrix's low-rank structure. Under this assumption, the $q$ latent factors exert a comparable total influence on the $p$ codified features in $\bX$, implying that the ratio $p/q$, our benchmark dimensionality, quantifies the average number of codified features influenced by each latent factor.
  
\end{remark}

\begin{remark}
In Assumption \ref{asu:3}, we allow the number of latent factors, $q$, to grow at a polynomial rate with respect to $n$ and $p$. Specifically, $q$ may increase slightly slower than $n$ and $p^{1/2}$. While most existing theoretical results on latent factor models \citep[e.g.,][]{zhang2020note, chen2019joint} and GVA methods for Poisson data \citep{hall2011theory} assume a fixed $q$, our setting permits a diverging number of latent factors. This relaxation is particularly relevant in EHR studies, where an increasing number of codified features in $\bX$ often reflects more complex and comprehensive disease conditions -- naturally requiring a larger latent space for accurate representation.
However, theoretical analysis under the regime $q \to \infty$ is significantly more challenging than in the fixed-$q$ case. The main technical difficulty arises from the term $\bV_j^\top(\bB \bUvec_Y + \bW)$ inside the exponential link, which complicates the control of the associated stochastic processes. We address this challenge using tools from high-dimensional empirical process theory. 
\end{remark}

\begin{remark}
It is not hard to show that when $q\rightarrow\infty$ under Assumption \ref{asu:3} and all entries in the coefficient matrix $\bB_{0}$ are generated independently from the standard normal distribution, both the condition $\bB_0 \in  \Omega_B$ in Assumption \ref{asu:2} (ii) and the SNR condition $\psi^2 > C_{\psi}$ used in Lemma \ref{lem:lin:cont} will hold with the probability approaching $1$ for any constant $C_{\psi}>0$. This illustrates our assumptions on $\bB_{0}$ are reasonable.  
\end{remark}

We start from presenting the error rate for the supervised estimator $\widetilde\btheta$.
\begin{theorem}[Preliminary supervised estimation]
\label{thm:sup:conv}
Under Assumptions \ref{asu:1}--\ref{asu:3}, we have 
\[
\Err(\widetilde\btheta)=\Op\left(\max\{\psi,1\}\Big(\frac{ q^{1/2}}{n^{1/2}}+\frac{ q^{2-\epsilon}}{p^{1-\epsilon}}\Big)\right).
\]
\end{theorem}
The first term $\max\{\psi,1\}{q^{1/2}}/{n^{1/2}}$ of $\Err(\widetilde\btheta)$ comes from the original empirical log-likelihood $\ell_{L}(\btheta)$. The second term $\max\{\psi,1\}{q^{2-\epsilon}}/{p^{1-\epsilon}}$ reflects the approximation error introduced by the GVA and is approximately of order $p^{-1}$ when $q$ is small. This behavior is conceptually aligned with the convergence rate established for GVA in repeated measurement models with $q = 1$ \citep{hall2011theory}. 

Compared with \cite{chen2019joint}, where $q$ is fixed, $\bV$ is fully estimated, $\bW_i$'s are treated as fixed parameters with no covariates, our supervised rate is roughly proportional to $n^{-1/2} + p^{-1}$, potentially faster than their $n^{-1/2} + p^{-1/2}$ rate when $p = o(n)$ due to borrowing knowledge from external pre-trained embeddings.
In the EHR setting, where the feature dimension $p$ is large and the labeled sample size $n$ is relatively small, the error $\Err(\widetilde{\btheta})$ is typically dominated by the component associated with limited labeled data. Consequently, the supervised estimator $\widetilde{\btheta}$ can suffer from substantial estimation error, limiting its reliability in high-dimensional, label-scarce environments. In contrast, the semi-supervised estimator $\widehat{\btheta}$ leverages both labeled and unlabeled data to improve estimation accuracy. We now derive its convergence rate and demonstrate how the additional information from unlabeled data mitigates the impact of small $n$. To this end, we first establish a key lemma characterizing the convergence behavior of the EM-GVA algorithm.

\begin{lemma}
[Linear contraction of EM-GVA] Recall that $\widehat\btheta\subt$ is the estimator after the $t$-th iteration in the EM algorithm. Under Assumptions \ref{asu:1}--\ref{asu:3}, there exists some constant $C_{\psi}>0$ such that when the SNR $\psi^2:=\bB_{Y,0}\trans\bLambda_0^{-1}\bB_{Y,0} > C_{\psi}$ and $\Err(\widehat\btheta\subt)=\op(\psi^{-1})$,
\[
\Err(\widehat\btheta\subtpone)\leq \{\nu+\op(1)\} \Err(\widehat\btheta\subt)+\Op\left(\frac{\psi q^{1/2}}{N^{1/2}}+\frac{\psi q^{2-\epsilon}}{p^{1-\epsilon}}\right),\quad t=0,1,\ldots,T,
\]
for some constant $\nu\in(0,1)$.
\label{lem:lin:cont}
\end{lemma}

Lemma \ref{lem:lin:cont} implies that $\widehat\btheta\subt$, obtained at each iteration $t$ of the EM-GVA algorithm, converges to the truth $\btheta_0$ with an error that decays linearly. 
The SNR condition $\psi^2 > C_{\psi}$ requires that the mean difference between $\bZ_i\mid Y_i=1$ and $\bZ_i\mid Y_i=0$ is sufficiently large compared to the variance of the random effect $\bW_i$. 
This condition is crucial for the contraction property of EM algorithm, and similar assumptions have been used in prior convergence analyses \citep{balakrishnan2017statistical, cai2019chime}.
The linear contraction process in Lemma \ref{lem:lin:cont} also requires the error of the initial estimator $\widehat\btheta\subzero=\widetilde\btheta$ to be moderately small.

By leveraging Theorem \ref{thm:sup:conv} and Lemma \ref{lem:lin:cont}, we establish the convergence rate of the SCORE estimator $\widehat\btheta$ in Theorem \ref{thm:semisup}. Specifically, Theorem \ref{thm:sup:conv} ensures the consistency of the initial estimator $\widetilde\btheta$ used to initialize the EM-GVA algorithm, while Lemma \ref{lem:lin:cont} shows that the estimation error of $\widetilde\btheta$ is further reduced through the iterative EM-GVA procedure.

\begin{theorem}[Convergence of SCORE]
\label{thm:semisup}
Under Assumptions \ref{asu:1}--\ref{asu:3}, there exist some constants $C_{\psi},D>0$ such that when $\psi^2 > C_{\psi}$ and $T>D\log(N/n)$,
\[
\Err(\widehat\btheta)=\Op\left(\frac{\psi q^{1/2}}{N^{1/2}}+\frac{\psi q^{2-\epsilon}}{p^{1-\epsilon}}\right).
\]
\end{theorem}
Compared with the supervised estimator $\widetilde\btheta$, $\widehat\btheta$ is still subject to the GVA error $\psi{q^{2-\epsilon}}/{p^{1-\epsilon}}$ but has a much faster convergence rate in its first part $\psi{q^{1/2}}/{N^{1/2}}$ as $N\gg n$. In fact, $N$ is more than a hundred times of $n$ in our EHR application example. It is also important to note that $\Err(\widehat\btheta)$ is free of the influence from the label size $n$ as long as $n$ is sufficient to ensure that $\Err(\widehat\btheta\subzero)=\Err(\widetilde\btheta)=\op(\psi^{-1})$, which is easy to justify by combining Assumption \ref{asu:3} and Theorem \ref{thm:sup:conv}. Theorem \ref{thm:semisup} also implies that our hybrid EM-GVA algorithm requires only an order of $\log(N/n)$ EM iterations to achieve the desirable convergence rate. Typically, this computational cost of exploiting the large unlabeled sample is moderate due to the logarithm.

Finally, we introduce Corollary \ref{coro:pred} for the convergence rate of the phenotyping probability $\gamma\supGVA\subnew$ and the embedding estimator $\widehat\bE\subnew$ for some new subject. 
\begin{corollary}[Quality of phenotyping and embedding for some new subject]
\label{coro:pred}
Under all assumptions and conditions in Theorem \ref{thm:semisup}, and assuming that $(\bx\subnew,\bu\subnew)$ is a realization of model (\ref{eq:model}) (parameterized by $\btheta_0$) independent of the training sample, we have
\begin{align*}
\gamma\supGVA\subnew-\Pr(Y=1|\bar\bxi=\bar\bxi\subnew,\bU=\bu\subnew)&=\Op\left(\frac{\psi^2 q^{1/2}}{N^{1/2}}+\frac{\psi q^{1/2-\epsilon}}{p^{1/2-\epsilon}}\right)=\op(1),\\
\big\|\widehat\bE\subnew -\bar\bxi\subnew\big\|_2 &=\Op\left(\frac{q^{1-\epsilon}}{p^{1/2-\epsilon}}\right)=\op(1),
\end{align*}
where $\gamma\supGVA\subnew = \gamma\supGVA(\bx\subnew,\bu\subnew;\widehat\btheta)$ is defined in (\ref{equ:def:pred}) and $\widehat\bE\subnew$ is defined in (\ref{equ:Enew}).
\end{corollary}
Recalling Proposition \ref{prop:1}, $\Pr(Y=1| \bar\bxi=\bar\bxi\subnew,\bU=\bu\subnew)$ has no information loss about $Y\subnew$ compared to $\Pr(Y=1| \bX=\bx\subnew, \bU=\bu\subnew)$. The convergence of $\gamma\supGVA\subnew$ to the former is due to the fact that our EM-GVA algorithm essentially recovers the latent embedding and uses it to impute the unobserved $Y$. Different from the error rate of $\widehat\btheta$, the $p$-driven error terms of $\gamma\supGVA\subnew$ and $\widehat\bE\subnew$ are proportional to $p^{1/2-\epsilon}$ but not $p^{1-\epsilon}$. This is because errors in the subject-level GVA term $\widehat{\bm{m}}\sy_i$ has first-order influence on the predicted probability for $Y$ but only the second-order influence on the model estimator $\widehat\btheta$.

\section{Simulation}

\subsection{Setup and Benchmark Methods}

We first evaluate the performance of SCORE estimators and compare it with existing methods with respect to both estimation and prediction accuracy using simulated datasets. For data generation, we set $\bV = c\widetilde\bV_{(,1:q)}$ where $\widetilde\bV$ is obtained by the eigen-decomposition  $\widetilde\bSigma=\widetilde\bV\widetilde\bLambda\widetilde\bV\trans$ on the auto-regressive matrix $\widetilde\bSigma= (0.5^{|i-j|})_{i,j\in[p]}$, and $c>0$ is specified such that $\sqrt{ q/p} \bV$ is orthonormal. We then set the true $\bLambda_0 = 4(0.1^{|i-j|})_{i,j\in[p]}$ and $\bB_0 = (0*\bone_q, 0.2*\bone_q, 0.8*\bone_q)$, and generate $\bW_i\sim \mathcal N(\bm 0_{q}, \bLambda_0)$, $U_i\sim Poisson(2)$, $Y_i \sim \expit(-0.2 + 0.5U_i)$,  $\bZ_i = \bV(\bB_0\bUvec_{iY} + \bW_i)$, and $\bX_i\sim Poisson(e^{\bZ_i})$ from the PALM model. We randomly choose $n$ samples to have observed $Y_i$, with the remaining unlabeled.  

To study the behavior of the optimization algorithm and assess SCORE's robustness to various model specifications, we vary the following generative parameters: (a) the labeled size $n=50, 100, 200, 400$; (b) the full sample size $N=1000, 2000, 5000, 10000$; (c) the number of candidate features $p=100,200,400, 800$; (d) the dimension of latent embedding $q=10, 20, 40, 80$; and (e) the specification of $\bUvec_{Y}$.  To simulate the presence of unobserved factors beyond $Y_i$ and $U_i$ that contribute to $\bX_i$, we introduce a setting in which the proposed model is slightly mis-specified. Specifically, we include an additional latent variable $C_i \sim \mathrm{Ber}(0.4)$ and define $\bZ_i =  \bV( \widetilde\bB \widetilde\bD_i+ \bW_i)$ with $\widetilde{\bD}_i = (1, U_i, Y_i, C_i)^\top$. We then construct two versions of the extended loading matrix $\widetilde{\bB}$ to reflect different signal strengths: (i) weak signal with $\widetilde \bB = (\bB_0, 0.2*\bone\trans_q)$; and (ii) strong signal with $\widetilde \bB = (\bB_0, 0.6*\bone\trans_q)$. (f) Subsequently, we generate $\bX_i$ from a zero-inflated Poisson distribution, with the zero-inflation probability set to either $0.05$ or $0.1$.

For comparison of parameter estimates, we considered both the supervised (sup) and semi-supervised (semisup) version of SCORE. The supervised SCORE is only fit on the small number of data with labels, whereas the semi-supervised SCORE utilizes both the labeled data and unlabeled data as proposed. For ablation studies, we also include the unsupervised version of SCORE without using the supervised estimator for initialization. We additionally considered two methods: (i) two-step SVD \citep{zhang2020note}, which was proposed for exploratory item factor analysis (IFA) for binary item response data, but here adopted for Poisson data, (ii) variational inference for probabilistic Poisson PCA model \citep{chiquet2018variational}. The difference between the model assumptions and optimization methods are summarized in Table \ref{tab:benchmark}. {We use the relative F-norm errors of the estimation of $\bB_0$ and $\bLambda_0$, defined as $err(\bBhat, \bB_0) =||\bBhat-\bB_0||_F/||\bB_0||_F$ and $err(\bLambdahat, \bLambda_0) = ||\bLambdahat-\bLambda_0||_F/||\bLambda_0||_F$. In addition, we evaluate the quality of the patient preresentation using cosine similarity of the predicted patient embedding with the true underlying patient embedding. We report the average classification performance of the predicted $\Pr(Y=1 \mid \bX, \bU)$ against the true outcome $Y$ based on the area under the receiver operating characteristics curve (AUC) and area under the precision-recall curve (PRAUC). All results are summarized based on $200$ simulations in each setting.}

\begin{table}[H]
    \footnotesize
    \centering
\begin{tabular}{l c c c c c c c}
\hline
\textbf{Method} & \textbf{$\bW$ random} & \textbf{known $\bV$} & \textbf{$\mathcal I_L$ used} & \textbf{$\mathcal I_U$ used} & \textbf{\# model params.} & \# \textbf{GVA params.}\\
\hline
Chique2018 & yes & no & yes & no  & $(r+2)p+pq$ & $nq$ \\
Zhang2020 & no & no & yes & no  & $(r+2)p+pq$ & $\slash$ \\
SCORE(sup) & yes & yes & yes & no  & $(r+2)q+q^2$ & $nq$\\
SCORE(unsup) & yes & yes & no & yes & $(r+2)q+q^2$ & $2(N-n)q$\\
SCORE(semisup) & yes & yes & yes & yes  &$(r+2)q+q^2$ &$2Nq$\\
\hline
\end{tabular}
\caption{Summary of the methods under comparison.}
\label{tab:benchmark}
\end{table}

\subsection{Results}

\paragraph{Parameter Estimation} 
We first investigate how relative F-norm errors of model parameters change with varying $N, n, p, q$.   As expected, with $p\gg q$, increasing the sample size of the labeled data ($n$) reduces the error for the supervised methods, while increasing the sample size of unlabeled data ($N$) only reduces the error for our semi-supervised SCORE (Figure \ref{fig:Nn}). Specifically, when $p=400$ and $q=20$, with $N=5000$, increasing $n$ from $50$ to $400$ decreases the relative error of $\bB$ from supervised SCORE from $0.72$ to $0.30$, while semi-supervised SCORE achieves an even lower error, decreasing from $0.12$ to $0.10$. In contrast, increasing $N$ from $1000$ to $10000$ for fixed $n=50$ improves semi-supervised SCORE performance, reducing error from $0.21$ to $0.08$, while supervised SCORE remains unchanged.

Figure \ref{fig:pq} shows the change in relative F-norm error with varying $p,q$ when $N=5000$ and $n=100$. We observe that as $q/p\to 0$, errors for supervised SCORE do not converge to $0$ due to the insufficient sample size, whereas errors for semi-supervised SCORE converge to $0$ as expected. Specifically, when $p=400$ and $q=10$, the relative error for estimation of $\bB$ from supervised SCORE remains at $0.58$, whereas the error for semi-supervised SCORE reduces to $0.08$. Additional results for unsupervised SCORE are shown in Appendix. The unsupervised SCORE shows high relative F-norm error across all scenarios compared with the semi-supervised SCORE due to the lack of convergence, highlighting the importance of initialization with the supervised estimator.

\paragraph{Patient Representations} 
Next, we investigate the mean cosine similarity between the true subject embedding $\bar\bxi_i$ and the estimated $\widehat\bE_i = \left\{(1-\widehat\gamma\supGVA_i)\widehat\bE_i\szero + \widehat\gamma\supGVA_i\widehat\bE_i\sone \right\}$ where $\widehat\bE_i\szero = \widehat\bB (1, U_i, 0)\trans + \widehat{\bm m}_i\szero$ and $\widehat\bE_i\sone = \widehat\bB (1,U_i, 1)\trans + \widehat{\bm m}_i\sone$. The results with varying $N$ and $n$ with $p=400$ and $q = 20$ are shown in Figure \ref{fig:sim_cos}. The patient-level embeddings generated with semi-supervised SCORE have the highest cosine similarities with the truth, outperforming its supervised counterparts and other methods. For instance, when $N=1000$ and $n=50$, the cosine similarity for semi-supervised SCORE reaches $0.91$, compared to $0.88$ for supervised SCORE and only $0.78$ for \cite{chiquet2018variational}. 

\paragraph{Classification performance}
In addition, we compare the classification performance of our method against competing approaches
Figure \ref{fig:sim_AUC} presents the mean AUC and mean PRAUC, along with standard deviations from $200$ simulations for each method, evaluating their robustness under various generative model specifications.
We can observe that (a) As expected, increasing the size of labels improves predictive performance for supervised and semi-supervised methods. (b) Unsurprisingly, our semi-supervised SCORE method singularly benefits from an increase in the size of unlabeled set. (c) Our method is more advantageous when $p$ is large. (d) Our method performs well even when $q$ is large, provided that $q$ remains small relative to $p$. (e) The performance of our method deteriorates with stronger misspecification, but it still outperforms the other methods. (f) As the proportion of zero inflation increases, modeling the counts as a Poisson distribution becomes slightly less advantageous.

\begin{figure}[H]
    \centering

    \includegraphics[width=16cm]{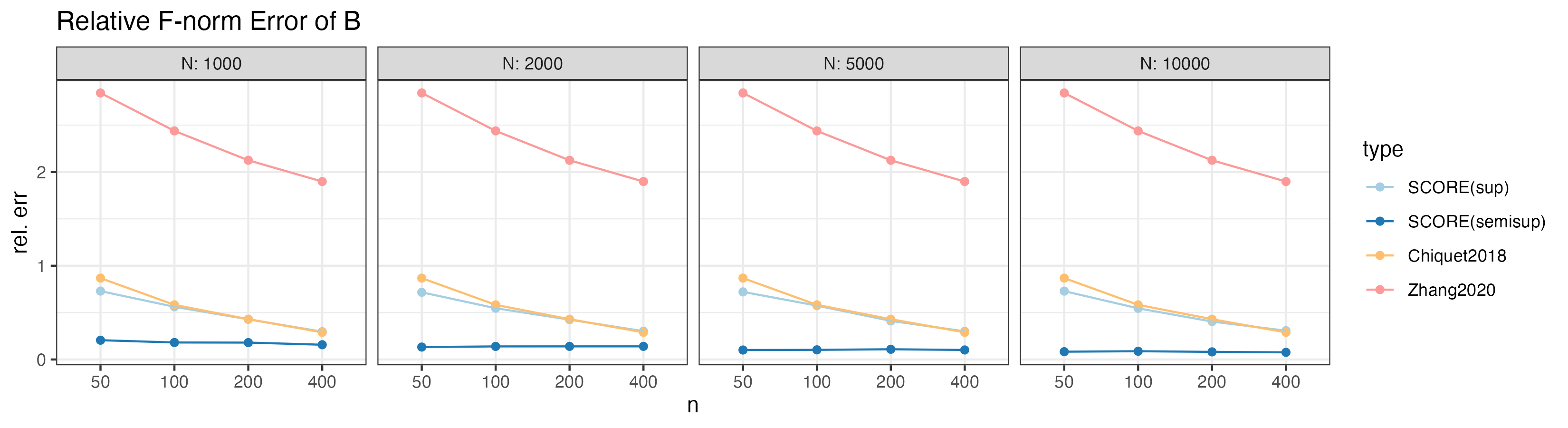} 
    \includegraphics[width=16cm]{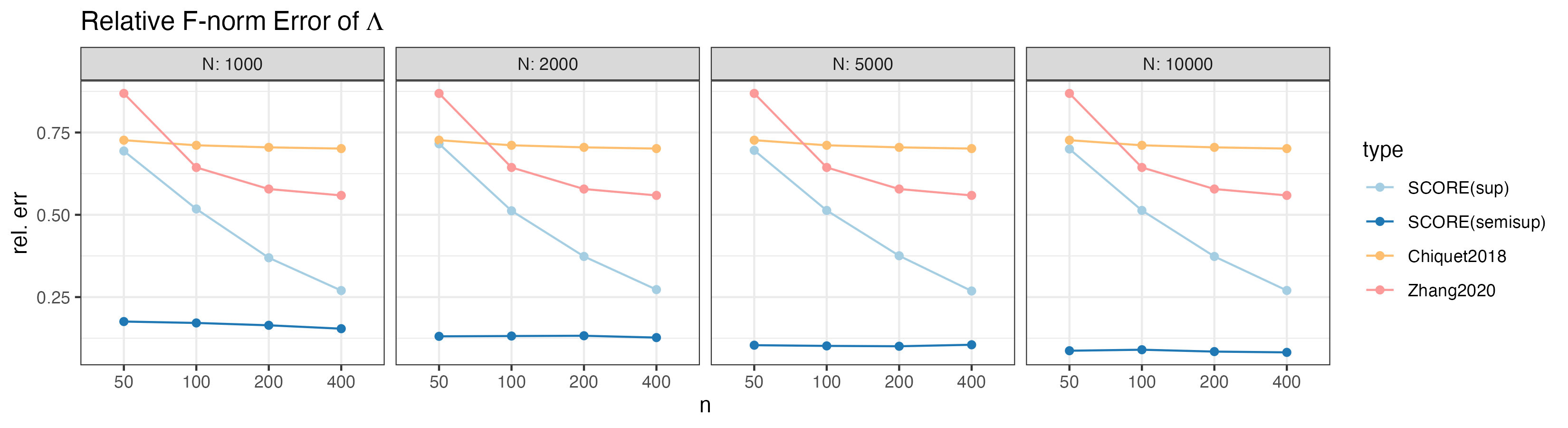}
    \caption{
    Relative F-norm errors in estimating $\bB$ and $\bLambda$ across different methods. Shown are the relative errors $\|\widehat\bB - \bB_0\|_F/\|\bB_0\|_F$ and $\|\widehat\bLambda - \bLambda_0\|_F/\|\bLambda_0\|_F$ under varying values of $N$ and $n$, with $p = 400$ and $q = 20$ held fixed.}
    \label{fig:Nn}
\end{figure}

\begin{figure}[H]
    \centering
    \includegraphics[width=16cm]{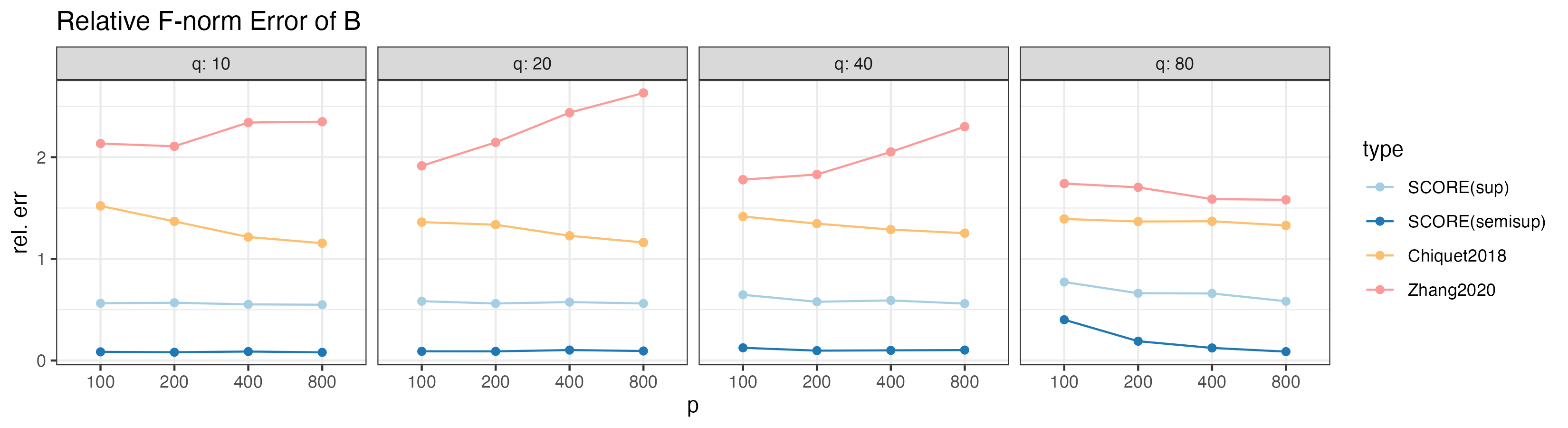}
    \includegraphics[width=16cm]{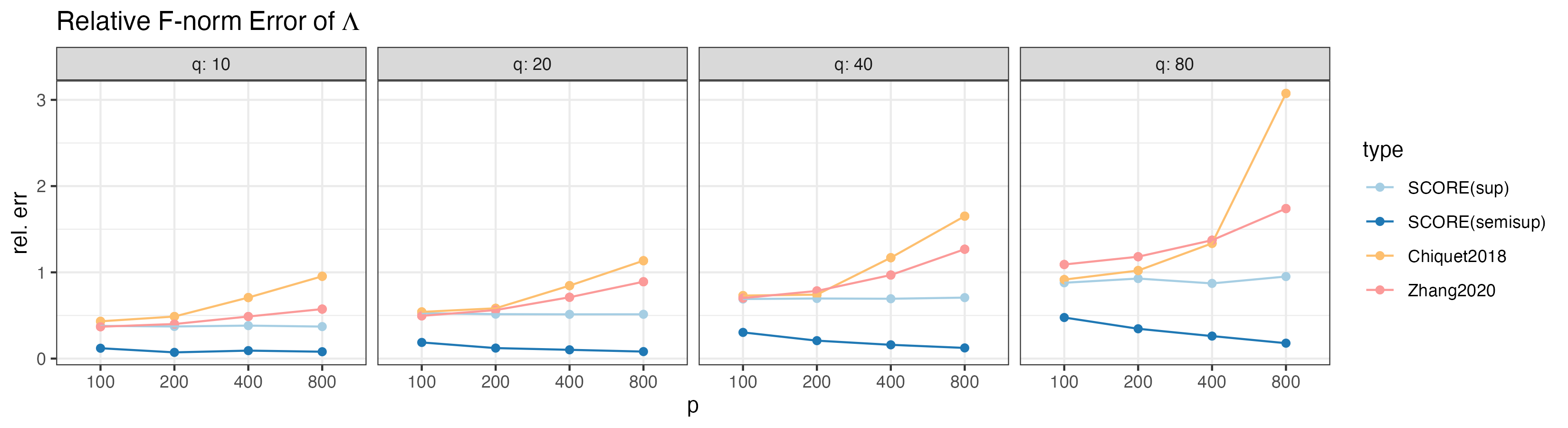}
    \caption{
    Relative F-norm errors in estimating $\bB$ and $\bLambda$ across different methods. Shown are the relative errors $\|\widehat\bB - \bB_0\|_F/\|\bB_0\|_F$ and $\|\widehat\bLambda - \bLambda_0\|_F/\|\bLambda_0\|_F$ under varying values of $p$ and $q$, with $N=5000$ and $n=100$ held fixed.}
    \label{fig:pq}
\end{figure}

\begin{figure}[H]
    \centering
    \includegraphics[width=16cm]{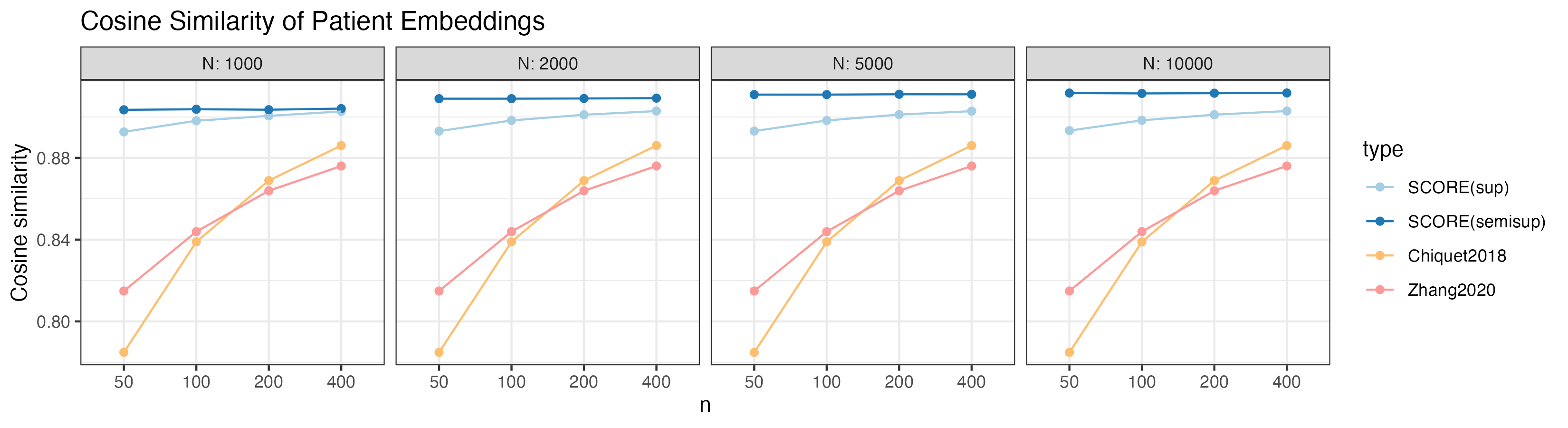}
    \caption{Mean cosine similarity between true and predicted patient embeddings across different methods. Results are shown for varying values of $n$ and $N$, with $p = 400$ and $q = 20$ held fixed.}
    \label{fig:sim_cos}
\end{figure}

\begin{figure}[H]
    \centering
    \includegraphics[width=17cm]{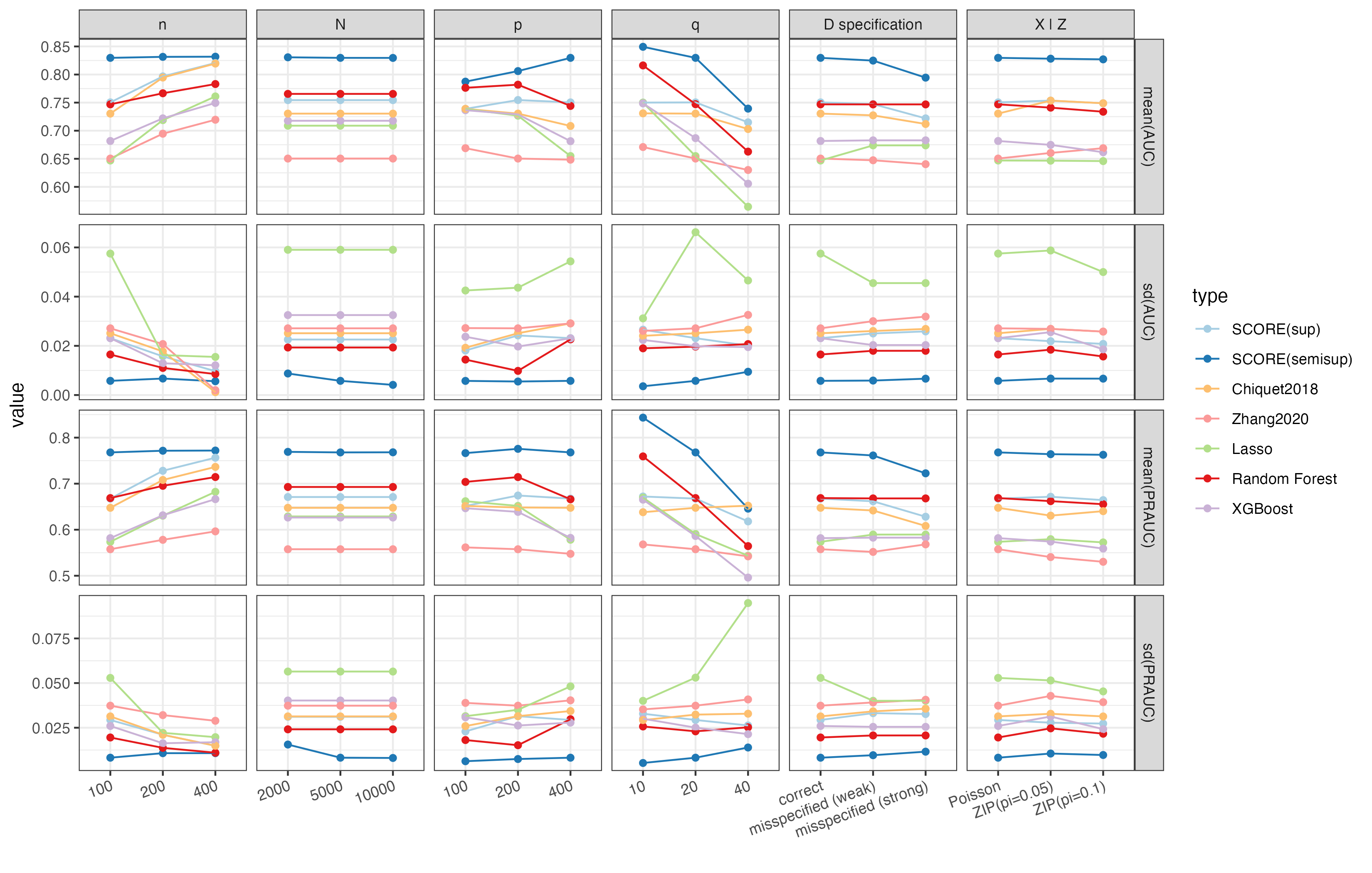}
    \caption{
    Classification performance of different methods under various generative settings. When one generative parameter is varied, the others are fixed at $n=100$, $N=5000$, $p=400$, and $q=20$ with correct model specifications.}
    \label{fig:sim_AUC}
\end{figure}

\section{Phenotyping of Disability for MS Patients}\label{sec:realdata}
We applied the proposed method to phenotype disability status in 16,091 patients with MS using EHR data from Mass General Brigham (MGB). The dataset includes both codified data, such as diagnoses, medications, and procedures, and narrative clinical notes, which were converted into structured clinical concept identifiers (CUIs) via natural language processing (NLP).

Disability is a key indicator of MS progression and its impact on daily life, making it essential for timely intervention and effective management. However, because disability measures are not routinely captured in EHRs, their utility for research and clinical decision-making is limited.
To address the absence of routinely recorded disability measures in EHRs, we leverage a disease registry linked to EHR data to infer disability status from the rich clinical information captured during routine care. The registry includes two validated indicators of disability: the clinician-assessed Expanded Disability Status Scale (EDSS), ranging from 0 to 10, and the patient-reported Patient-Determined Disease Steps (PDDS), ranging from 0 to 8. Patients with EDSS $\ge$ 5 or PDDS $\ge$ 4 were categorized as having a high level of disability ($Y = 1$); otherwise, they were labeled as having a low level of disability ($Y = 0$).
Our objective was to use EHR data to phenotype binary disability status at each clinical visit, using PDDS as the primary guide. By leveraging both structured variables and NLP-derived features from clinical notes, we aimed to identify patient clusters that not only capture meaningful distinctions in PDDS-defined disability but also correspond strongly with the alternative EDSS measure which is more commonly used in clinical trials \citep{van2017outcome}. Furthermore, we sought to learn patient embeddings from the algorithm that are predictive of future disability progression, enabling clinically useful stratification beyond the labeled data.

We included EHR features relevant to the MS diagnosis or disability using the online narrative and codified feature search engine \citep{xiong2023knowledge}. For each visit, we aggregate feature counts within a 6-month window before and after the visit to construct the feature matrix $\bX_{once}$. In addition, we manually selected five features relevant to disability $\bX_{man}$: \emph{wheelchair}, \emph{weakness}, \emph{mental health}, \emph{dalfampridine} and \emph{speech and language disorder} to assist in the prediction of disability status. We then obtained the embeddings of all selected concepts $\bX = (\bX_{man}, \bX_{once})$ through multi-source representation learning which leverages information from both EHR data as well as contextual embeddings for the textual descriptions of the clinical concepts \citep{xiong2023knowledge}. We applied SVD to the feature-level embedding matrix and got $\widetilde\bV\in\mathbb R^{p\times q}$ as the first $q$ left singular vectors. For our model, we specified $\bV = [(\bm I_5,\bm 0)\trans, \widetilde\bV]$.

We compared the performance of our method with other existing methods for the classification of disability status based on the EDSS and PDDS scores. We considered five methods as benchmarks: (1) Lasso, (2) Random Forest, (3) XGBoost, (4) supervised fully connected neural network (\textit{NN(sup)}), and (5) variational autoencoder (\textit{NN(VAE)}). For Lasso, random forest and XGBoost, the only output is the predicted score. For the two neural network based models, we obtained both the predicted score and the low-dimensional patient embeddings for downstream tasks. We trained each model using $n=50, 100, 150, 200$ labeled samples, and validated the models based on $n_{test} = 500$ labeled samples. Figure \ref{fig:MGB_MS}A shows the performance of each model with varying training sizes. The semi-supervised SCORE method achieves the highest AUC and PRAUC, along with the lowest Brier score, particularly when the labeled sample size is small ($n = 50, 100$). As the labeled sample size increases, all methods show comparable performance. Notably, SCORE consistently yields the lowest standard errors across all metrics by effectively leveraging the large pool of unlabeled data during training.

We further evaluated whether models trained using PDDS-based labels could generalize to predicting EDSS-defined disability status. Notably, our method outperformed alternative approaches in this cross-outcome evaluation (Figure \ref{fig:MGB_MS}B), indicating stronger generalizability and an improved ability to capture the true underlying disability status across both patient-reported and clinician-assessed scales. A key strength of our approach lies in its ability to generate low-dimensional patient embeddings that encode clinically meaningful information. To assess their utility beyond the initial classification task, we used embeddings trained on current PDDS-based disability status to predict future disability progression. Specifically, we trained a logistic regression model to predict disability status two years later, using baseline embeddings (or predicted scores for methods that do not produce embeddings). Cross-validation results, shown in Figure \ref{fig:MGB_MS}C, demonstrate that our method achieved the highest AUC and PRAUC and the lowest Brier score, highlighting its robustness and value for downstream tasks such as long-term patient stratification.

\begin{figure}[h!]
    \centering
\includegraphics[width=16cm]{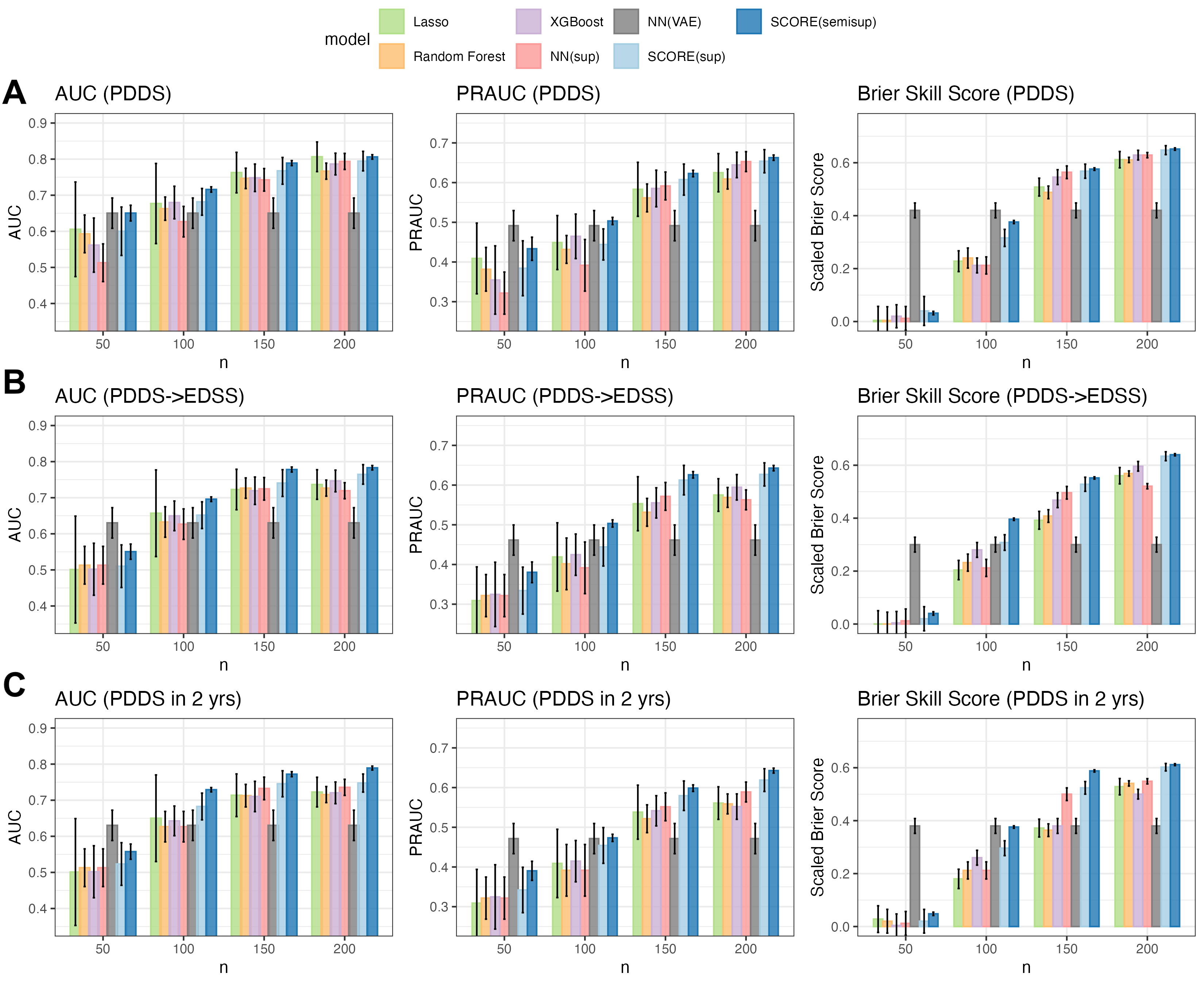}
\caption{Classification performance with varying $n$ for phenotyping of disability for MS patients at MGB. \textit{PDDS:} trained and validated using disability status based on PDDS, \textit{PDDS-EDSS:} trained using disability status based on PDDS and validated on EDSS-based labels.}
\label{fig:MGB_MS}
\end{figure}

\section{Discussion}
This paper introduces SCORE, a semi-supervised representation learning framework that models high-dimensional EHR data using a novel Poisson-adapted latent factor mixture model and pre-trained code embeddings to generate informative patient phenotypes. 
By combining a hybrid EM-GVA algorithm with theoretical guarantees, SCORE leverages unlabeled data to improve prediction accuracy and embedding quality, outperforming existing methods in both simulations and real-world applications.

{
Theoretical results on low-rank Poisson models have been limited, often under the restrictive assumption that the rank is either one \citep{hall2011asymptotic, hall2011theory} or fixed \citep{zhang2020note, chen2019joint}. In contrast, we establish the error rate of our EM-GVA algorithm considering a diverging number of latent factors, a scenario more suited to the complexities of EHR data. Our theoretical analysis not only addresses a critical gap in VI for low-rank Poisson models but also provides insights on the GVA error analysis in other contexts, which is necessary due to its potential flaws including slow optimization convergence and poor capture of the true posterior in certain scenarios \citep{yao2018yes}.
Moreover, by modeling the generative process $\bX \mid Y$, SCORE is robust to label noises, allowing observed $Y_i$ to serve as a noisy proxy for latent cluster membership, as seen in conditions like MS where PDDS and EDSS provide complementary but imperfect views of disability severity. The clustering-based framework permits flexible initialization, where any consistent supervised estimator suffices without affecting the final error rate, enhancing generalizability across clinical outcomes.

Although SCORE is developed within the Poisson distribution framework, the framework is generalizable to other distributions such as negative binomial or zero-inflated distributions \citep{batardiere2024zero}, which may offer improved flexibility for handling over-dispersed data. The current SCORE algorithm focuses on a single binary outcome $Y$, it is also feasible to extend the methodology to support multiple outcomes and a wider range of clinical outcomes, including continuous and categorical variables, to broaden its applicability across diverse medical settings. 

In our real EHR application, we refined the construction of $\bV$ to incorporate several EHR concepts known, based on domain knowledge, to be strongly associated with the disability outcome of interest. This adjustment illustrates the importance of modeling both low-rank and sparse structures, as similarly addressed in the Factor Augmented Sparse Throughput (FAST) algorithm \cite{fan2024factor}. When influential features are known a priori, our framework flexibly accommodates additional factors in $\bV$ to represent these strong signals. In contrast, when such features are unknown, introducing a sparse mean component in the latent layer offers a promising strategy for adaptively capturing underlying sparse structure. Developing a more principled approach to incorporating these sparse components may further enhance the model's ability to identify and leverage key predictors or surrogate features. 
Moreover, when the true and external factor structures diverge substantially, especially in heterogeneous populations, methods that go beyond rescaling eigenvalues may be needed to capture the latent structure more fully. Together, these directions suggest a rich landscape for extending the current model to address a wider variety of data types and clinical contexts. 
}

\bibliographystyle{plainnat}
\bibliography{ref}

\end{document}